\documentclass[12pt]{article}
\usepackage{amsmath}
\usepackage{amssymb}
\usepackage{graphicx}
\usepackage{adjustbox}
\usepackage{natbib}
\usepackage{url} 
\usepackage{hyperref}
\usepackage{pdflscape}   
\usepackage{todonotes}
\usepackage{booktabs}
\usepackage{listings}
\usepackage{comment}
\usepackage{xcolor}
\definecolor{codegreen}{rgb}{0,0.6,0}
\definecolor{codegray}{rgb}{0.5,0.5,0.5}
\definecolor{codepurple}{rgb}{0.58,0,0.82}
\definecolor{backcolour}{rgb}{0.95,0.95,0.92}
\lstdefinestyle{mystyle}{
    backgroundcolor=\color{backcolour},   
    commentstyle=\color{codegreen},
    keywordstyle=\color{black},
    numberstyle=\tiny\color{codegray},
    stringstyle=\color{codepurple},
    basicstyle=\ttfamily\footnotesize,
    breakatwhitespace=false,         
    breaklines=true,                 
    captionpos=b,                    
    keepspaces=true,                 
    numbers=left,                    
    numbersep=5pt,                  
    showspaces=false,                
    showstringspaces=false,
    showtabs=false,                  
    tabsize=2
}
\usepackage{tabularx}   
\usepackage{array}      
\usepackage{multirow}
\usepackage{subcaption}

\usepackage{etoolbox}
\AtBeginEnvironment{table}{\renewcommand{\baselinestretch}{2}}
\AtEndEnvironment{table}{\renewcommand{\baselinestretch}{2}}
\AtBeginEnvironment{figure}{\renewcommand{\baselinestretch}{2}}
\AtEndEnvironment{figure}{\renewcommand{\baselinestretch}{2}}

\usepackage{tikz}
\newcommand*\circled[1]{\tikz[baseline=(char.base)]{
            \node[shape=circle,draw,inner sep=2pt] (char) {#1};}}

\usepackage{placeins}

\usepackage{multirow}

\usepackage[toc,page]{appendix}

\usepackage{comment}

\newcommand{\blind}{0}

\newcommand{\xiran}[1]{\textcolor{black}{#1}}
\newcommand{\sameh}[1]{\textcolor{black}{#1}}
\newcommand{\javi}[1]{\textcolor{black}{#1}}

\begin{document}

\def\spacingset#1{\renewcommand{\baselinestretch}%
{#1}\small\normalsize} \spacingset{1}

\if0\blind
{
  \title{\bf RCOMPSs: A Task-based Parallel Programming Model in R}
  \author{Xiran Zhang$^1$, Javier Conejero$^2$, Sameh Abdulah$^1$, Jorge Ejarque$^3$, Ying Sun$^1$\\ Rosa M. Badia$^2$, 
  David E. Keyes$^2$, 
  and Marc G. Genton$^1$ \\
  \small $^1$Computer, Electrical and Mathematical Sciences and Engineering Division, \\
  \small King Abdullah University of Science and Technology, Thuwal, KSA.\\
  \small $^2$Barcelona Supercomputing Center, Barcelona, Spain. \\
  \small $^3$Seqera Labs, Barcelona, Spain.
    }    
  \date{}
  \maketitle
} \fi

\if1\blind
{
  \bigskip
  \bigskip
  \bigskip
  \begin{center}
    {\LARGE\bf RCOMPSs: A Task-based Parallel Programming Model for R}
\end{center}
  \medskip
} \fi
\bigskip


\begin{abstract}

\xiran{R is widely used for scientific and statistical computing, and its package ecosystem supports a broad range of reproducible analyses. This paper introduces \textit{RCOMPSs}, an R binding to the \textit{COMPSs} runtime for task-based parallel execution. Users write standard R code and mark selected functions as tasks. The runtime then tracks dependencies, schedules tasks, moves data, and executes them across available resources. \textit{RCOMPSs} is intended to complement existing R parallel frameworks by supporting workflows that benefit from explicit task dependencies. We evaluate it with three representative algorithms, $k$-nearest neighbors classification, $k$-means clustering, and linear regression with predictions, on KAUST's Shaheen~III and BSC's MareNostrum~5. The experiments examine execution-time scaling and workflow trade-offs in single-node comparisons and distributed-memory runs. Overall, \textit{RCOMPSs} adds \textit{COMPSs} task-based execution to the R parallel ecosystem for scalable statistical workflows.}

\end{abstract}

\noindent%
\vfill

\newpage
\spacingset{1.75} 


\section{Introduction}
\label{sec:Introduction}




Statistical computing is critical for extracting insights from data, enabling researchers and analysts to model uncertainty, test hypotheses, and make data-driven decisions. Among the available tools, the R programming language is a powerful and widely used platform for statistical analysis and data science~\citep{R-base}. Its extensive ecosystem of packages, designed for statistical modeling, machine learning, data visualization, and reporting, makes R suitable for handling complex analytical tasks~\citep{Wickham2017}. Furthermore, its focus on reproducibility and active community support has established it as a core tool in data analytics, statistical research, and academic use.
\xiran{R predates today's multicore and manycore systems, but its ecosystem has grown with the needs of scientific computing. Python has followed a similar path: libraries and runtimes have extended the language's original execution model to use modern parallel hardware.}

Traditional sequential processing in R becomes a bottleneck as data volumes increase, particularly when handling large datasets or computationally intensive tasks. Parallel programming addresses this challenge by enabling concurrent execution, which significantly reduces computation time and improves efficiency, especially in workflows involving iterative algorithms, simulations, or large-scale transformations. By leveraging multicore and manycore architectures, R users can scale their analyses and extract insights from big data more effectively. Consequently, parallelism has become a vital enabler in extending its high-performance data analytics capabilities. 

R was initially designed as a single-threaded, memory-bound, interpreted language, which limits its scalability on high-performance computing (HPC) platforms~\citep{Stojanovic2016}. Although performance-critical components are often implemented in compiled languages such as C, C++, or Fortran, R still faces challenges, including limited native support for parallelism, high memory overhead from copy-on-modify semantics, and weak integration with distributed-memory systems~\citep{RcppParallel}. These limitations restrict its ability to fully exploit modern HPC infrastructures without substantial low-level optimization or external parallel frameworks. In the literature, several efforts, e.g., ExaGeoStat \citep{abdulah2018exageostat,abdulah2023large}, have  demonstrated how highly optimized C/C++ backends can accelerate domain-specific statistical workloads. However, these implementations target specific problems rather than offering general-purpose solutions.

In HPC environments, a runtime system is a software layer that can bridge the gap between a program and the hardware it runs on. Rather than requiring the user to explicitly manage \xiran{processes, data movement,} or processor assignments, the runtime automatically coordinates how tasks are executed, how data dependencies are resolved, and how computational resources are used. In this way, runtime systems make parallel processing more accessible by allowing researchers to focus on the statistical or analytical logic of their applications, while abstracting away the technical details of low-level parallelism. Well-known examples include OpenMP~\citep{chandra2001parallel}, StarPU~\citep{augonnet2009starpu}, PaRSEC~\citep{hoque2017dynamic}, and Kokkos~\citep{edwards2014kokkos}. 

COMP Superscalar (\textit{COMPSs})~\citep{badia_comp_2015, Web:COMPSs} is one of these advanced task-based runtime systems that automatically builds a task dependency graph at runtime and schedules tasks asynchronously based on data availability~\citep{badia_comp_2015}. It simplifies the work of programmers by allowing them to write code in a sequential style, while the runtime system transparently manages the parallel execution. Furthermore, \textit{COMPSs} is designed to scale from desktops to large HPC clusters, making it suitable for a wide range of data-intensive and scientific applications.

This work presents \textit{RCOMPSs}, a runtime system that extends the capabilities of the R language by enabling efficient task-based parallel execution on multicore and manycore systems. Built on top of the \textit{COMPSs} framework, \textit{RCOMPSs} introduces a new binding for R, allowing developers to write sequential R code while transparently benefiting from parallel execution. By annotating functions as tasks, users can offload execution management to the underlying runtime, which dynamically constructs a task dependency graph, schedules tasks across available resources, and handles data movement and synchronization. \textit{RCOMPSs} also supports fault tolerance through the built-in mechanisms of the \textit{COMPSs} runtime, including automatic task resubmission and exception management. A key aspect of \textit{RCOMPSs} is its infrastructure-unaware programming model. It allows users to develop applications once and run them seamlessly on different backends, \xiran{including single-node and distributed-memory execution settings}, without modifying the original implementation. \xiran{The runtime manages scheduling, dependency resolution, and data movement, while developers remain responsible for task decomposition and application design.} We demonstrate the effectiveness of \textit{RCOMPSs} through three representative statistical applications, \textit{k}-means clustering, \textit{k}-nearest neighbors (KNN) classification, and linear regression with prediction, evaluated \xiran{in single-node and distributed-memory settings}.
\xiran{In our experiments, \textit{RCOMPSs} reduces execution time relative to sequential R code for the workloads we tested, while also incurring the initialization and serialization costs of task-based execution. \textit{RCOMPSs} is a lower-level, runtime-oriented interface. Developers define task boundaries and data dependencies; the \textit{COMPSs} runtime handles scheduling, data movement, and synchronization. Higher-level statistical functions or workflow frameworks can be built on this interface, but \textit{RCOMPSs} does not decide how tasks should be designed or how statistical parameters should be chosen.}




This article is organized as follows: Section~\ref{sec:Related_work} reviews the current state-of-the-art capability of the R language for exploiting multicore and manycore CPU systems, focusing on current challenges and existing solutions; Section~\ref{sec:background} introduces task-based runtimes and the \textit{COMPSs} framework; Section~\ref{sec:RCOMPSs} presents \textit{RCOMPSs}, detailing its architecture, programming model, and functionalities; Section~\ref{sec:Application_kmeans} overviews the three example implementations developed in this work and evaluates the performance and behavior of \textit{RCOMPSs} on two supercomputers; Section~\ref{sec:comparison_with_other_packages} offers a comparative study of \textit{RCOMPSs} and three widely used R parallel packages; and Section~\ref{sec:Conclusions} summarizes the main findings and outlines directions for future work.


\section{Parallel Computing in R}
\label{sec:Related_work}


Parallel computing in R has advanced significantly through a diverse ecosystem of packages and frameworks that support \xiran{single-node and distributed-memory execution}. 
At the low level, interfaces to OpenMP and Intel Threading Building Blocks (TBB) are available via packages such as \textit{RcppParallel}~\citep{RcppParallel} to enable fine-grained control and high-performance parallelism within compiled code. For distributed-memory systems, \textit{Rmpi}~\citep{Rmpi} provides a low-level interface to the Message Passing Interface (MPI) for scalable execution on multiple nodes. While powerful, these low-level interfaces require users to manage communication, synchronization, and workload distribution explicitly, increasing complexity and reducing portability.  At a higher level of abstraction, R also offers user-friendly tools such as \textit{parallel}~\citep{R-base}, \textit{foreach}~\citep{foreach}, \xiran{\textit{doParallel}~\citep{doParallel}, which registers the \textit{parallel} backend for \textit{foreach},} and \xiran{the legacy \textit{snow}~\citep{snow} package, whose functionality has largely been superseded by \textit{parallel}}, which simplify \xiran{split-apply-combine computations and independent task execution} for multicore systems. These packages provide essential primitives for parallel execution, but still require users to manually handle data partitioning and synchronization, limiting scalability and usability for complex workflows or heterogeneous distributed environments.

More recently, the \textit{future} package~\citep{future} presents a flexible and unified framework for asynchronous and parallel evaluation across local and remote resources. Its design separates developer intent (what to parallelize) from user execution (how and where), making it well-suited for reproducible statistical workflows. Earlier R mechanisms also separated these concerns: \textit{snow}/\textit{parallel} exposed cluster selection through functions such as \texttt{setDefaultCluster()}, while \textit{foreach} backends used registration functions such as \texttt{registerDoNnn()}~\citep{R-base,snow,foreach}. \xiran{The package is best viewed as a foundation on which higher-level interfaces can be built; for example, load balancing and chunking choices are typically exposed through packages in the future ecosystem, such as \textit{future.apply}~\citep{future.apply} and \textit{furrr}~\citep{furrr}. The newer \textit{mirai} package~\citep{mirai} provides another lightweight framework for asynchronous and parallel computation in R. The \textit{largescaler} package is also related, with a focus on scaling R analyses to larger data and compute settings~\citep{largescaler}.} While integrations such as \textit{future.batchtools}~\citep{future.batchtools} enable HPC usage, the framework remains a high-level abstraction that depends on backend capabilities and system schedulers, such as the Simple Linux Utility for Resource Management (SLURM), a widely used HPC workload manager~\citep{yoo2003slurm}.

In the context of large-scale data analytics, R can interoperate with big data platforms such as Hadoop and Apache Spark through packages like \textit{sparklyr}~\citep{sparklyr} and \textit{SparkR}~\citep{venkataraman2016sparkr}, which allow users to offload computation to distributed data engines. However, these integrations often come with limitations, such as restricted support for complex statistical methods, dependence on the underlying big data framework, and additional overhead in data movement and serialization. Comprehensive reviews of parallel computing approaches in R can be found in \cite{schmidberger2009state} and \cite{eddelbuettel2021parallel}.


\section{Task-based Runtimes and the COMPSs Framework}
\label{sec:background}


In HPC environments, task-based runtime systems play an essential role in bridging high-level programming models with the underlying hardware processing units~\citep{sterling2017survey}. They can abstract the task scheduling, data movement, synchronization, and fault tolerance from developers. This enables developers to express parallelism at a higher level, while relying on the runtime to manage efficient execution on heterogeneous platforms.
Modern task-based runtime systems such as StarPU~\citep{augonnet2009starpu}, PaRSEC~\citep{hoque2017dynamic}, Kokkos~\citep{edwards2014kokkos}, and \textit{COMPSs}~\citep{badia_comp_2015, Web:COMPSs} can dynamically construct and schedule dependency graphs, exploit parallelism across CPUs and GPUs, and adapt to distributed-memory environments. 

\textit{COMPSs} is a powerful task-based runtime system designed to simplify the parallelization of sequential applications with minimal effort and their automatic execution in \xiran{single-node and distributed-memory computing environments}. It addresses the complexities inherent in harnessing the power of multicore processors and large clusters by abstracting away low-level details such as task scheduling, data distribution, and resource management. At its core, \textit{COMPSs} operates on a task-based paradigm where applications are decomposed into independent units of work (tasks) linked by their data dependencies. This paradigm is underpinned by the \textit{COMPSs} runtime system, which orchestrates the execution of these tasks across the underlying infrastructure. The runtime functionality extends beyond mere task scheduling. It carefully analyzes the application code to identify data dependencies between tasks, constructing a dynamic task dependency graph that informs the scheduling decisions. This graph enables efficient parallelization and optimization of execution flow, as well as ensuring the validity of the execution results. Furthermore, the runtime handles crucial aspects of data management, ensuring data consistency by controlling access from multiple tasks, preventing race conditions, and maintaining data integrity.

Another key strength of \textit{COMPSs} is its robust infrastructure abstraction layer, which hides the complexities of distributed computing. The runtime is capable of handling a wide range of systems, from laptops to HPC clusters and supercomputers, within a single job request. It also incorporates pluggable connectors that facilitate seamless integration with various cloud platforms. This abstraction promotes portability across infrastructures. It is even capable of dynamically managing resource allocation based on application demands, scaling computing resources up or down (if the underlying infrastructure allows) to optimize performance while minimizing costs.

The \textit{COMPSs} runtime, implemented in Java, provides native support for Java applications and also offers interfaces for other programming languages such as C/C++ and Python (\textit{PyCOMPSs}~\citep{tejedor2017pycompss}), fostering flexibility and adaptability. By abstracting the complexities of parallel programming, \textit{COMPSs} enables developers to build robust and scalable distributed applications while its runtime orchestrates tasks, data, and resources across heterogeneous infrastructures. This design allows researchers to concentrate on algorithmic innovation rather than low-level execution details. Notably, \textit{PyCOMPSs} has been successfully applied in diverse domains, including engineering~\citep{amela2018executing}, earth sciences~\citep{nmmb-monarch}, life sciences~\citep{pymdsetup}, and quantum computing~\citep{rosnet}, demonstrating its broad applicability and impact. 


\section{RCOMPSs Runtime System}
\label{sec:RCOMPSs}



Starting from pure R code, where each line is executed sequentially, we propose a programming model that can detect and exploit the inherent parallelism hidden within the sequential code. This idea follows the \textit{COMPSs} programming model, aiming to ease the parallelization of sequential applications with minimal impact on them. This paradigm offers several key benefits, including simplified development efforts, concealing distributed execution complexity, ensuring infrastructure independence, and providing application portability and scalability.

In particular, this work proposes the \textit{RCOMPSs} programming model, which builds upon the \textit{COMPSs} runtime and is designed to streamline the parallelization of R code. \textit{RCOMPSs} adopts an intuitive sequential programming model, allowing users to write applications as standard R scripts while focusing on two key aspects: (i) identifying functions suitable for asynchronous parallel execution, and (ii) annotating these functions as tasks using a simple R call. The underlying \textit{COMPSs} runtime system then transparently manages the parallel execution. It automatically detects data dependencies between tasks, constructs a directed acyclic graph (DAG) of task dependencies, and dynamically schedules tasks across available resources in distributed environments such as HPC systems, clusters, or clouds. By abstracting away low-level parallelism concerns, \textit{RCOMPSs} enables efficient and scalable execution with minimal changes to the original R code. \xiran{\textit{RCOMPSs} and \textit{targets}~\citep{targets} both use dependency-aware graphs, but they address different layers of the workflow. \textit{targets} focuses on reproducible R analysis pipelines, caching, and invalidation of pipeline steps, whereas \textit{RCOMPSs} exposes \textit{COMPSs} task execution, scheduling, data movement, and resource management for HPC environments. This design allows \textit{RCOMPSs} to provide integrated task execution, scheduling, data movement, and resource management for complex DAG-based workloads on distributed HPC infrastructures.}


Although the \textit{COMPSs} runtime provides critical functionality, it lacks language-specific features, so \textit{RCOMPSs} must address many challenges. For example, \textit{RCOMPSs} needs to provide a syntax and mechanism to define tasks in R code, it also needs to provide a mechanism to swap objects transparently to the application, it must deal with object serialization, it must communicate with the \textit{COMPSs} runtime following its API, and it must provide the R code that receives the task execution requests and execute them (capability to execute the isolated task code). In this section, we elaborate on these challenges and how they have been addressed from an integration perspective, from high-level R syntax (API) to low-level technical details.

\subsection{RCOMPSs Architecture}
\label{subsec:architecture}

\begin{figure}[t]
    \centering
    \subfloat[]{
        \includegraphics[width=0.315\textwidth]{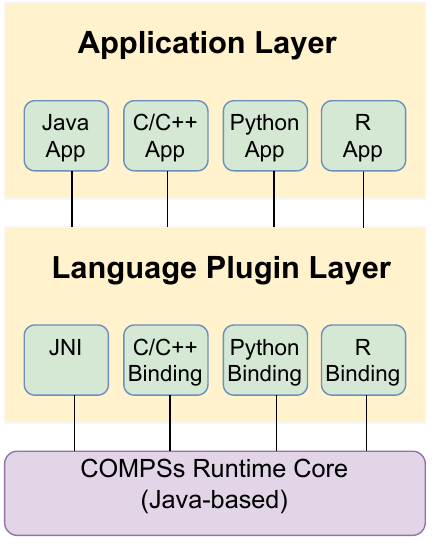}
        \label{fig:compss}
    }
    \hspace{12mm}
    \subfloat[]{
        \includegraphics[width=0.35\textwidth]{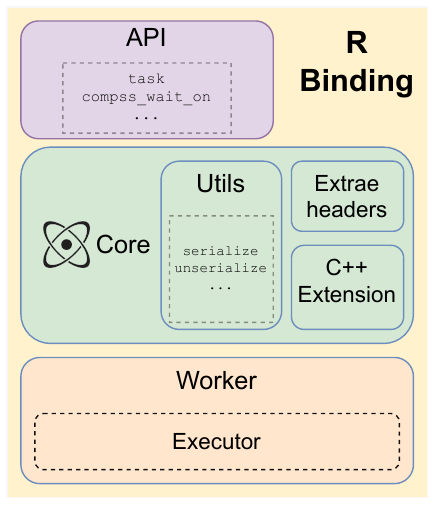}
        \label{fig:r_binding_architecture}
    }
    \caption{Architecture of (a) the \textit{COMPSs} runtime and its language bindings and (b) the \textit{RCOMPSs} binding, including the R API, core utilities, C++ extension, and worker and executor modules.}
    \label{fig:compss_rcompss}
\end{figure}

\textit{COMPSs} supports various popular programming languages, including Java (natively), Python (\textit{PyCOMPSs}), and C/C++ through bindings, leveraging existing knowledge and promoting code reusability. \textit{PyCOMPSs}, in particular, provides a set of decorators to define functions as tasks and an API to synchronize data within Python. It uses persistent Python worker processes to minimize task invocation overhead, prevent repeated interpreter launches, and enhance execution throughput. Thus, \textit{PyCOMPSs} has inspired the support of the R language. Figure~\ref{fig:compss} illustrates the \textit{COMPSs} architecture and the various programming languages it supports, including Python and R.

\textit{RCOMPSs} is integrated into the \textit{COMPSs} framework as a new binding, extending language support to R. It has been developed with \textit{COMPSs} versions 3.3 to 3.3.3. \xiran{\textit{RCOMPSs} is primarily supported on Linux-based systems. It can also be used on macOS and on Windows 11 through WSL2, provided that the required \textit{COMPSs}, Java, R, and system dependencies are available.} 
The extension introduces new dependencies, mainly the R interpreter and required libraries. The \textit{COMPSs} installer has been updated to optionally include \textit{RCOMPSs} and automatically detect or install missing R packages, simplifying the setup process.
The current supported Java versions are 17--21; Java~11 was used for the experiments reported here with \textit{COMPSs}~3.3.2.
The installation also checks for MPI header files to configure the Extrae tracing integration. This is an installation requirement of the \textit{COMPSs} and Extrae environment rather than an indication that \textit{RCOMPSs} uses MPI as its user-level execution model; the R bindings delegate task scheduling, worker management, and data movement to the \textit{COMPSs} runtime.

Support for R code is enabled by the development of a new \textit{COMPSs} binding, allowing R functions within R applications to be intercepted, registered with the \textit{COMPSs} runtime, and executed in distributed environments. To achieve this, the \textit{RCOMPSs} package is built on top of the \textit{COMPSs} binding commons interface (language plugin layer), as illustrated in Figure~\ref{fig:compss}. \textit{RCOMPSs} is designed to enable R users to benefit from automatic parallelization and execution provided by the task-based programming model while preserving data dependencies to ensure the correctness of the results. \textit{RCOMPSs} leverages the internal architecture of the \textit{COMPSs} runtime system, inheriting key functionalities such as task scheduling, data dependency analysis, data transfer, and fault tolerance at no additional cost. To enable this seamless integration, \textit{RCOMPSs} includes several features specifically designed to support interoperability with \textit{COMPSs}.

Figure~\ref{fig:r_binding_architecture} illustrates the \textit{RCOMPSs} architecture, which is adapted to the R programming environment. It is primarily composed of three modules: \emph{API}, \emph{Core}, and \emph{Worker}. The top layer, \emph{API}, provides the interface to R applications and defines the \textit{RCOMPSs} programming model syntax. Since \textit{COMPSs} follows a task-based paradigm, the \emph{API} provides minimal yet essential functionality for defining tasks and managing data synchronization. It is also extensible, allowing future integration of additional \textit{COMPSs} features. 

Below the \emph{API}, there is the \textit{RCOMPSs} \emph{Core} module. It is the heart of the \textit{RCOMPSs} programming model. It is responsible for performing all necessary actions to be taken for task preparation (parameter serialization, task registry, and object tracking -- most of them implemented within the \emph{Utils} submodule) and \textit{COMPSs} requests for execution or data retrieval (implemented within the \emph{C++ Extension} submodule). The \emph{Core} module also implements an Extrae headers submodule, enabling integration with Extrae~\citep{Web:EXTRAE}, a dynamic instrumentation tool designed for parallel applications based on shared memory (e.g., OpenMP, pthreads), message passing interface (MPI), or hybrid programming models, to generate performance traces. 

Finally, the \textit{RCOMPSs} binding contains a \emph{Worker} module used by the \textit{COMPSs} runtime. The runtime launches one worker process per node, and each worker manages one executor process per core on that node. Thus, a worker comprises several executors, and the total process counts depend on the number of nodes and cores allocated on each platform. Each executor remains active throughout the application execution time and receives requests from the \textit{COMPSs} runtime to execute each particular task using the specified parameters.

This architecture has been successfully integrated into the \textit{COMPSs} framework repository and has been demonstrated to be effective in introducing a task-based programming model to the R language.

\subsection{Technical Details}
\label{subsec:technical_details}

This subsection describes the key components of the \textit{RCOMPSs} API and packaging strategy, highlighting how they enable parallel task execution, efficient data handling, and system interoperability.

\subsubsection{Runtime Binding via C++ Interface}

\textit{COMPSs} provides the \emph{binding-commons} API natively. It is a C++ library intended for bindings to communicate with the \textit{COMPSs} runtime. Consequently, \textit{RCOMPSs} implements a C++ connector capable of invoking \emph{binding-commons} API functions. The \emph{binding-commons} API provides low-level functions for requesting runtime start/stop, executing a task (which requires task information and parameters), and data synchronization. These are minimal functions to be implemented by any binding to interact with \textit{COMPSs} and are implemented in the \emph{C++ Extension} module of \textit{RCOMPSs}. However, \emph{binding-commons} also provides more functions that can be used to extend the binding functionalities (e.g., define files as parameters, task constraints, task groups, checkpointing, etc.). 

To enable communication between the R layer and the underlying \textit{COMPSs} runtime, \textit{RCOMPSs} leverages \textit{Rcpp} package, a useful tool for integrating C++ code with R~\citep{Rcpp}. The C++ connector in \textit{RCOMPSs}, built using \textit{Rcpp}, invokes functions from the \emph{binding-commons} API whenever a task-related operation is called from the R interface (e.g., \verb|compss_start()| or \verb|task()|). This design abstracts the complexity of runtime communication, allowing R users to interact with the \textit{COMPSs} runtime without writing C++ code. \textit{Rcpp} ensures efficient data exchange between R and C++ while maintaining the performance and flexibility needed for high-performance computing workflows.

\subsubsection{Worker Deployment and Parallel Execution in R}

For worker deployment on \xiran{single-node and distributed-memory systems}, \textit{COMPSs} launches the \textit{RCOMPSs} worker processes described in Section~\ref{subsec:architecture}; each worker manages persistent R executor processes on its node.  These executors handle successive task requests, avoiding a new R interpreter launch for every task. Future versions may investigate NUMA-aware task placement and shared-memory optimizations to reduce memory transfers and improve locality when data reuse is high~\citep{amela2018executing}.
\subsubsection{Object Management}

The \emph{binding-commons} API function for task execution in \textit{COMPSs} requires metadata from the task definition, including the number of parameters, their types, data direction (e.g., input/output), and the actual parameter values. To maintain language-agnostic interoperability, \textit{COMPSs} bindings use a file-based mechanism for parameter passing. Each parameter must be serialized into a file before task submission, allowing the runtime to handle data transfer across distributed systems if necessary. These files are then deserialized at the target location to reconstruct the original objects, ensuring consistency and compatibility.

\subsection{RCOMPSs Syntax}
\label{subsec:syntax}

The \textit{RCOMPSs} programming model provides five core interface functions for task-based parallelization (Table~\ref{tab:rcompsssyntax}).

\begin{table}[h!]
    \caption{User-facing \textit{RCOMPSs} API functions.}
    \centering
    \begin{tabular}{|l|p{0.6\linewidth}|}
        \hline
        \textbf{Function} & \textbf{Description} \\
        \hline
        \texttt{compss\_start()} & Initializes the \textit{COMPSs} runtime system. \\
        \hline
        \texttt{task()} & Defines an individual \textit{RCOMPSs} task. \\
        \hline
        \texttt{compss\_barrier()} & Synchronizes execution by waiting for all previously submitted tasks to complete. \\
        \hline
        \texttt{compss\_wait\_on()} & Waits for and retrieves the result of a specific object generated by any task. \\
        \hline
        \texttt{compss\_stop()} & Terminates the \textit{COMPSs} runtime system. \\
        \hline
    \end{tabular}

    \label{tab:rcompsssyntax}
\end{table}

These functions assist R application developers in parallelizing their sequential code. The \verb|task()| function is the most important; it allows developers to define functions as tasks, which are then executed in parallel, with input and output data managed by the \textit{COMPSs} runtime.


After defining tasks using \verb|task()| to construct the workflow, \verb|compss_wait_on()| is invoked to retrieve the results produced by any task when they are required in the main program. Otherwise, data are automatically transferred between tasks without the need for explicit synchronization. This function blocks execution until the corresponding task completes, at which point the runtime returns the resulting value.

\verb|compss_start()| and \verb|compss_stop()| are aimed at initializing and terminating the \textit{COMPSs} runtime, especially in interactive R environments. Finally, \verb|compss_barrier()| synchronizes execution by waiting for all preceding tasks to finish.



\begin{figure}[h!]
    \centering
    \begin{minipage}[t]{0.66\linewidth}
        \vspace{0pt} 
\begin{lstlisting}[
  language=R,
  basicstyle=\ttfamily\fontsize{10}{7.5}\selectfont
]
library(RCOMPSs)
source("task.R")
compss_start()
get_mean.dec <- task(get_mean, "task.R",
                info_only = F, return_value = T)
get_sd.dec <- task(get_sd, "task.R",
                info_only = F, return_value = T)
standardize.dec <- task(standardize, "task.R",
                info_only = F, return_value = T)
# Define three tasks below
mu <- get_mean.dec(X)
sigma <- get_sd.dec(X)
X_standardized <- standardize.dec(X, mu, sigma)
# Get the result
X_standardized <- compss_wait_on(X_standardized)
compss_stop()
\end{lstlisting}
    \end{minipage}
    \begin{minipage}[t]{0.33\linewidth}
        \vspace{-20pt} 
        \centering
        \includegraphics[width=\linewidth,height=280pt]{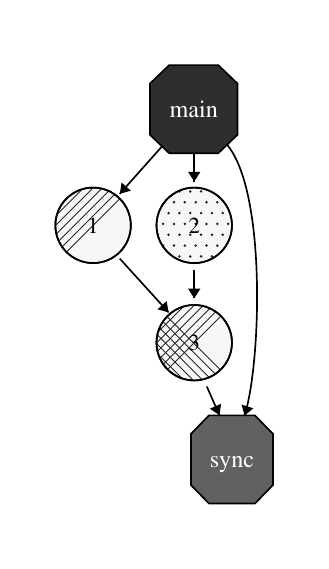}
    \end{minipage}
\caption{Vector standardization in \textit{RCOMPSs}: driver code in \texttt{job.R} (left) and the resulting task DAG (right). The mean and standard-deviation tasks can run concurrently; standardization depends on both results.}
    \label{fig:toy_example}
\end{figure}

To demonstrate the usage of \textit{RCOMPSs}, we present a simple example that standardizes a numeric vector. This example requires two files: one named \verb|task.R|, which defines three functions:

\begin{lstlisting}[language=R]
get_mean <- function(x) mean(x) # returns the mean of x
get_sd <- function(x) sd(x) # returns the standard deviation of x
standardize <- function(x, mu, sigma) (x - mu) / sigma # standardization
\end{lstlisting}

The other file, \verb|job.R|, contains the code that launches the execution of \textit{RCOMPSs}, as shown in Figure \ref{fig:toy_example}. \xiran{This separation is needed because the \textit{COMPSs} runtime reads task definitions separately from the main program. Unlike Python, R does not provide a direct equivalent of \texttt{if \_\_name\_\_ == "\_\_main\_\_"} to prevent top-level code from being executed when a script is sourced. Keeping task definitions in \texttt{task.R} and the driver code in \texttt{job.R} therefore avoids unintended execution while allowing the runtime to load the task functions.}

\definecolor{standardization_blue}{HTML}{27E7F5}
\definecolor{standardization_brown}{HTML}{8B0000}
\definecolor{standardization_yellow}{HTML}{F5F527}
\definecolor{standardization_green}{HTML}{28FC9A}
\definecolor{standardization_red}{HTML}{ff0000}

The runtime automatically generates the directed acyclic graph (DAG) shown in Figure~\ref{fig:toy_example}. During execution of the code above, \textit{RCOMPSs} creates an asynchronous task for each call to a function annotated with \verb|.dec|, which identifies the function as a \textit{Task}. It then constructs a DAG representing the resulting tasks and their data dependencies. In the DAG, the \texttt{main} node corresponds to \verb|compss_start|, while nodes \circled{1}, \circled{2}, and \circled{3} represent the tasks that compute \texttt{mu}, \texttt{sigma}, and \texttt{X\_standardized}, respectively. As illustrated by the task graph, the \texttt{mu} and \texttt{sigma} tasks have no dependency on one another and can therefore execute in parallel. In contrast, the \texttt{X\_standardized} task depends on the outputs of both tasks and can begin only after they have completed. The \texttt{sync} node corresponds to \verb|compss_wait_on|, after which the workflow concludes with \verb|compss_stop|. This example demonstrates that developers can express the application using a sequential programming style, while \textit{RCOMPSs} automatically identifies task dependencies and manages the available parallelism at runtime.

\section{Benchmarking Applications}
\label{sec:Application_kmeans}



In this study, we implement three parallel algorithms using \textit{RCOMPSs} and evaluate their performance on \xiran{single-node and distributed-memory systems}. The primary contribution of this paper is to enable task-based parallelization in R. To demonstrate its effectiveness, we benchmark widely used algorithms: \textit{k}-means clustering, \textit{k}-nearest neighbors (KNN) classification, and linear regression with predictions. The parallel implementations follow algorithmic patterns adapted from~\cite{dislib}.  The analysis of the three applications follows the same workflow and obtains very comparable results. Therefore, we discuss only one of the three applications in the main paper: \textit{k}-means clustering. Detailed information on the other two can be found in the Supplementary Materials.


\subsection{\textit{k}-means Clustering}

\textit{k}-means is a widely used unsupervised learning algorithm that partitions a dataset into \( k \) clusters by minimizing within-cluster variation~\citep{hartigan1979algorithm}. Given a dataset \( \{\mathbf{x}_1, \mathbf{x}_2, \dots, \mathbf{x}_n\} \subset \mathbb{R}^d \), the goal is to assign each data point to the cluster with the nearest centroid, the mean position of all points in a cluster, representing its geometric center in the feature space, thereby grouping similar points and keeping clusters as compact as possible. Formally, \textit{k}-means seeks to minimize the within-cluster sum of squares (WCSS):
\[
\underset{C}{\text{arg min}} \sum_{i=1}^k \sum_{\mathbf{x}_j \in C_i} \|\mathbf{x}_j - \boldsymbol{\mu}_i\|^2,
\]
where \( \boldsymbol{\mu}_i = \frac{1}{|C_i|} \sum_{\mathbf{x}_j \in C_i} \mathbf{x}_j \) is the centroid of cluster \( C_i \).

In \textit{RCOMPSs}, the parallel \textit{k}-means algorithm is implemented using a task-based parallel model that divides the dataset into fragments and processes them concurrently. The algorithm begins by initializing \( k \) cluster centroids. Data fragments are generated independently using the \texttt{fill\_fragment} task using random numbers, as the synthetic data is generated on the fly and not read from files. \xiran{For stochastic tasks, reproducibility is controlled explicitly by the user. In the data-generation step, each fragment receives a deterministic task-specific seed, for example \texttt{iseed = seed + i}, so concurrently executed tasks do not depend on execution order for their random-number streams. The same explicit-seeding strategy was used in the baseline implementations: the \textit{parallel} version used \texttt{mclapply()} with \texttt{set.seed(seed + a)} inside each worker call; the \textit{future} version created each future with \texttt{set.seed(seed + f)} and \texttt{seed = NULL}; and the \textit{furrr} version used \texttt{future\_map()} with \texttt{set.seed(seed + .x)} and \texttt{furrr\_options(seed = NULL)}.} In each iteration, \texttt{partial\_sum} tasks are executed in parallel to compute local sum of points and counts per cluster within each fragment. These local results are then combined using a hierarchical tree of \texttt{merge} tasks to compute the updated global centroids. \xiran{Thus, this example is not purely embarrassingly parallel: the parallel fragment computations are followed by dependent merge and centroid-update stages, and each iteration depends on the centroids produced by the previous one.}

\begin{figure}[!htb]
    \centering
    \includegraphics[width=0.4\linewidth]{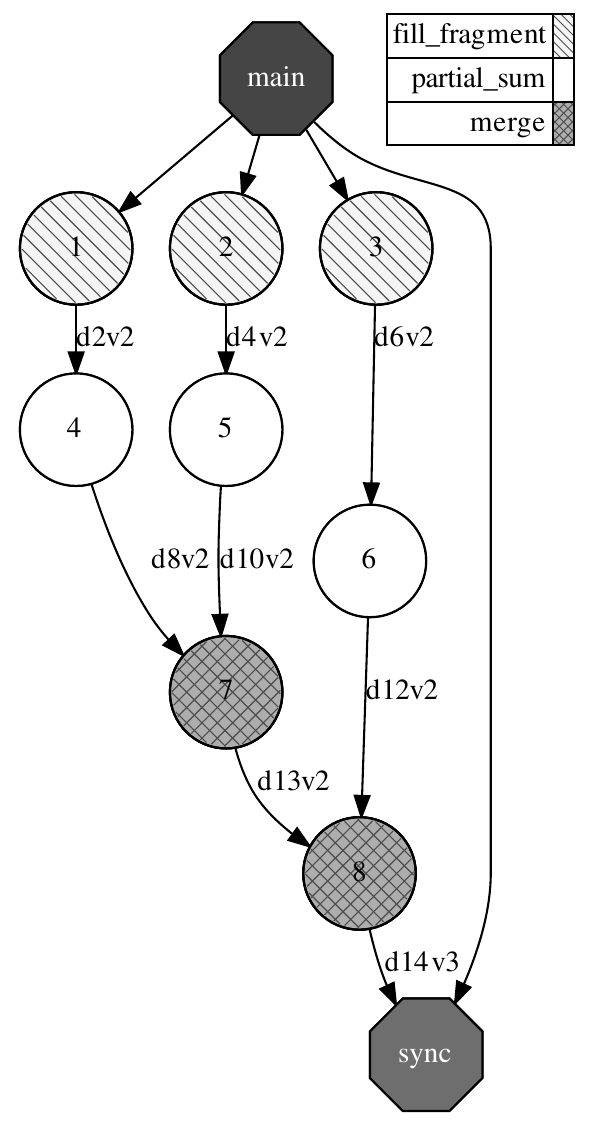}
    \caption{Task DAG for one iteration of parallel \textit{k}-means. Diagonally hatched circles create data fragments, unfilled circles compute local cluster sums and counts, and cross-hatched circles merge those results. Arrows show data dependencies; the dark octagon marks synchronization.}
    \label{fig:DAG-Kmeans}
\end{figure}

Convergence is determined using the \texttt{converged} function, which compares centroids across iterations, and the process continues until the centroids stabilize. Figure~\ref{fig:DAG-Kmeans} illustrates the parallel \textit{k}-means DAG for a small case with only one iteration, where tasks for partial computation, merging, and synchronization are organized based on their data dependencies. \xiran{The \texttt{dXvY} notation denotes data identifier X and version Y. It records the data dependency between tasks and constrains their execution order; versioning lets the runtime distinguish successive data states and expose independent tasks for parallel execution. The number of branches in the example follows from the data fragments and the hierarchical merge structure, so the graph need not be interpreted branch by branch. In terms of the WCSS objective, \texttt{fill\_fragment} creates the data blocks, \texttt{partial\_sum} assigns points in each block to their nearest current centroid and accumulates local per-cluster sums and counts, and \texttt{merge} combines these local sufficient statistics to form the centroid updates used in the next iteration.} Diagonally hatched circles represent \texttt{fill\_fragment} tasks, unfilled circles represent \texttt{partial\_sum} tasks, cross-hatched circles represent \texttt{merge} tasks, and the final dark octagon (\texttt{sync}) denotes synchronization at the end of the iteration.


\subsection{Experimental Testbed}
Our experiments were conducted on two CPU-based systems: KAUST Shaheen~III, consisting of $4{,}608$ dual-socket nodes, each equipped with 96-core AMD EPYC Genoa processors running at 2.4 GHz and 384 GB of memory, and the MareNostrum~5 (MN5) system at the Barcelona Supercomputing Center (BSC), consisting of $6{,}408$ dual-socket nodes with 56-core Intel Sapphire Rapids 8480+ processors running at 2.0 GHz and 256 GB of memory.

Our experiments are based on \textit{RCOMPSs} v1.0, built using \textit{COMPSs} v3.3.2. 
The runtime environment includes Java v11. The \textit{RCOMPSs} binding uses \textit{RMVL} v1.1.0.1~\citep{rmvl} and \textit{Rcpp} v1.0.14; the separate single-node comparison also includes an implementation using R's base \textit{parallel} package. These R packages are available through the Comprehensive R Archive Network (CRAN)\footnote{\url{https://cran.r-project.org/}}, while \textit{RCOMPSs} is publicly accessible at \url{https://github.com/stsds/rcompss}. The repository README documents the required software, installation and execution steps, and the scripts and input data needed to reproduce the experiments. \xiran{\textit{RCOMPSs} is not currently distributed through CRAN, so \texttt{install.packages("RCOMPSs")} is not the installation route at this stage.}
\xiran{Although \textit{RCOMPSs} is not currently distributed through CRAN, we plan to make the R package available through R-universe to simplify package installation when \textit{COMPSs} is provided by a system administrator.}


\subsection{\textit{k}-means Performance on \xiran{Single-Node and Distributed-Memory Systems}}


To evaluate \textit{RCOMPSs} on \xiran{single-node and distributed-memory systems}, we conduct single-node and multi-node experiments on Shaheen~III and MN5. On single nodes, we assess weak and strong scalability up to 128 executor cores on Shaheen~III and 80 executor cores on MN5, reserving the remaining cores for \xiran{runtime and master processes}. On distributed-memory configurations, we measure weak and strong scaling up to 32 nodes. In the main paper, we only detail the \textit{k}-means results, as KNN and linear regression exhibit similar qualitative behavior.

\begin{figure*}[t!]
    \centering
    
    \begin{subfigure}[b]{0.48\linewidth}
        \centering
        \includegraphics[width=\linewidth, page=3]{Figures/JCGS_scalability_plots/kmeans_scalability_shape.pdf}
        \caption{Single-node strong scaling}
        \label{fig:single_strong_kmeans}
    \end{subfigure}
        \hfill
    \begin{subfigure}[b]{0.48\linewidth}
        \centering
        \includegraphics[width=\linewidth, page=4]{Figures/JCGS_scalability_plots/kmeans_scalability_shape.pdf}
        \caption{Single-node weak scaling}
        \label{fig:single_weak_kmeans}
    \end{subfigure}
    \\
    \begin{subfigure}[b]{0.48\linewidth}
        \centering
        \includegraphics[width=\linewidth, page=1]{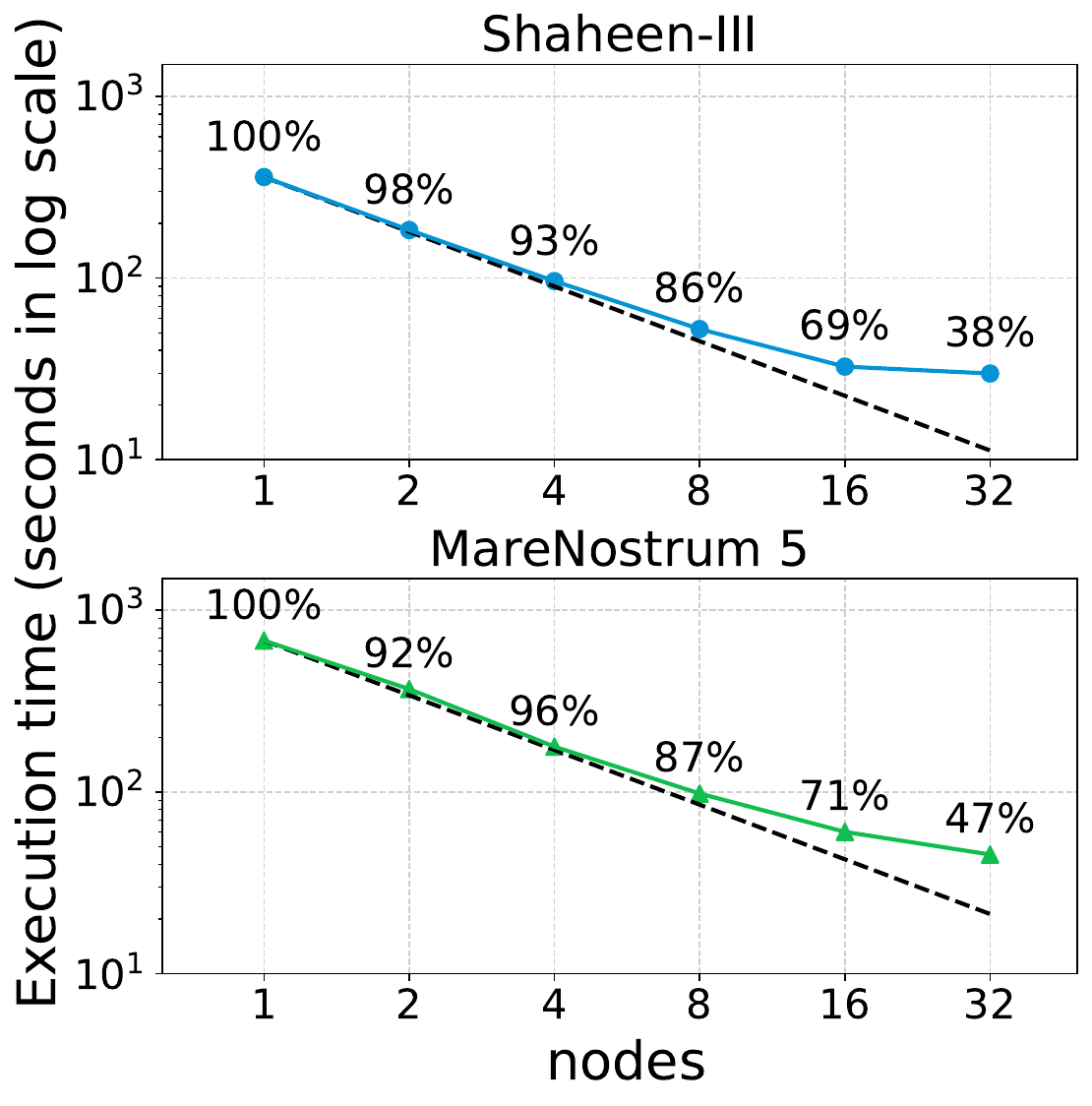}
        \caption{Multi-node strong scaling}
        \label{fig:multi_strong_kmeans}
    \end{subfigure}
    \hfill
    \begin{subfigure}[b]{0.48\linewidth}
        \centering
        \includegraphics[width=\linewidth, page=2]{Figures/JCGS_scalability_plots/kmeans_scalability_shape.pdf}
        \caption{Multi-node weak scaling}
        \label{fig:multi_weak_kmeans}
    \end{subfigure}

    \caption{Strong and weak scaling of \textit{k}-means on Shaheen~III and MareNostrum~5. Panels show execution time and parallel efficiency versus executor-core count for single-node runs and node count for multi-node runs (up to 32 nodes).}
    \label{fig:kmeans_scalability}
\end{figure*}

Figure~\ref{fig:kmeans_scalability} summarizes \textit{k}-means scalability across single-node and multi-node settings. On a single node, strong scaling (Figure~\ref{fig:single_strong_kmeans}) uses a 51,200,000 × 100 dataset: on Shaheen~III, \textit{k}-means achieves over 80\% parallel efficiency at 64 cores before declining due to synchronization and communication costs; on MN5, it scales to 80 cores with roughly 41\% efficiency. Single-node weak scaling (Figure~\ref{fig:single_weak_kmeans}) increases the dataset from 864,000 × 50 at 1 core to 110,592,000 × 50 at 128 cores: Shaheen~III scales efficiently up to 64 cores and retains over 60\% efficiency at 128 cores, whereas MN5 shows moderate scaling up to 48 cores and drops to about 43\% efficiency at 80 cores, likely due to task-scheduling and I/O overheads. On distributed-memory systems, strong scaling (Figure~\ref{fig:multi_strong_kmeans}) uses a 1,221,840,896 × 100 dataset and reaches 38\% efficiency on Shaheen~III and 47\% on MN5 at 32 nodes, with higher node counts increasingly affected by synchronization and I/O overheads. Distributed-memory weak scaling (Figure~\ref{fig:multi_weak_kmeans}) grows the dataset from 38,182,528 × 100 on 1 node to 1,221,840,896 × 100 on 32 nodes, with efficiency declining from 100\% to 61\% on Shaheen~III and remaining above 64\% on MN5 at 32 nodes.

The results for KNN and linear regression exhibit similar trends as shown in the Supplementary Materials. Overall, scalability tends to degrade at higher core/node counts for all algorithms across both systems.


\subsection{Execution Traces Analysis}

\begin{figure*}[h!]
    \centering
    \includegraphics[width=0.8\linewidth]{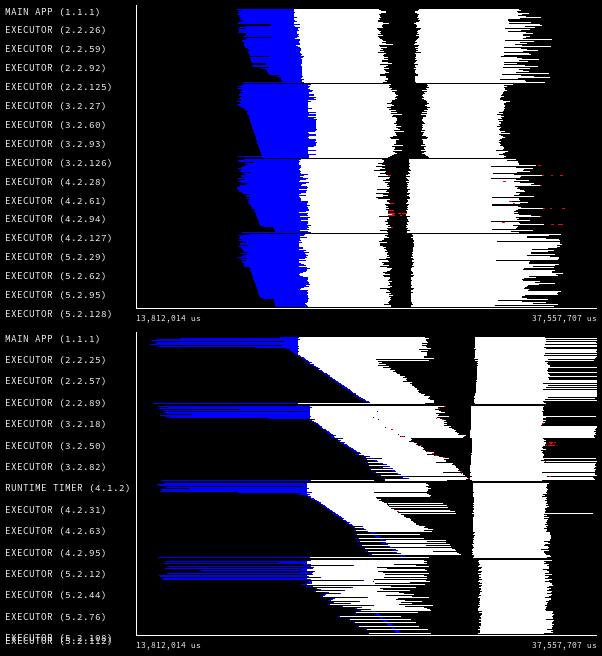}
    \caption{Execution traces for two \textit{k}-means iterations on $163{,}840{,}000 \times 5$ data points using four nodes: Shaheen~III (top) and MareNostrum~5 (bottom). Time is in microseconds. Lanes show the main application (MAIN APP), task executors (EXECUTOR), and runtime instrumentation (RUNTIME TIMER). Blue, white, and red blocks denote \texttt{fill\_fragment}, \texttt{partial\_sum}, and \texttt{merge} tasks, respectively.}
    \label{fig:traces-Kmeans}
\end{figure*}

To better understand the behavior of the three algorithms on the two target systems, we present execution traces generated using the Extrae profiling tool within \textit{RCOMPSs}. Figure~\ref{fig:traces-Kmeans} shows the execution trace of the parallel \textit{k}-means algorithm. The algorithm follows a map-reduce pattern with two iterations separated by a merge phase. \xiran{In each panel, time progresses from left to right. The rows labeled MAIN APP, EXECUTOR, and RUNTIME TIMER correspond to the driver process, worker-side executor processes, and runtime instrumentation, respectively. The first iteration consists of fragment-generation and \texttt{partial\_sum} tasks; the intervening \texttt{merge} task combines local per-cluster sums and counts to update centroids, after which the second iteration can start.} The initial blue blocks represent parallel \texttt{fill\_fragment} tasks, followed by white \texttt{partial\_sum} tasks used for centroid computation. On Shaheen~III, these tasks execute concurrently, and the red \texttt{merge} task must complete before the second iteration begins. This dependency creates a visible black gap between the two white regions in the trace, indicating synchronization overhead between rounds.
On MN5, worker initialization is noticeably slower, which leads to less parallelism during the blue \texttt{fill\_fragment} phase, i.e., often appearing partially sequential. This delay propagates, negatively affecting the scheduling and execution of the white \texttt{partial\_sum} and red \texttt{merge} tasks. As a result, the overall execution time on MN5 is significantly higher compared to Shaheen~III.


\section{\javi{Design Choices and Comparison with Existing Parallel Frameworks}}
\label{sec:comparison_with_other_packages}

\javi{
\textit{RCOMPSs} complements the existing R parallel ecosystem. Packages such as \textit{future}, \textit{furrr}, and \textit{parallel} provide flexible ways to run independent computations, including the process-based configurations used in our single-node comparison. \textit{RCOMPSs} delegates task scheduling, dependency management, and data movement to the \textit{COMPSs} runtime, allowing the same task-based application to run across nodes. This approach adds initialization and serialization costs while reducing the need to manage dependencies and inter-node data movement in application code.
}

\begin{figure}[h!] 
\centering 
\begin{subfigure}{0.315\textwidth} 
    \centering 
    \includegraphics[width=\linewidth,page=1]{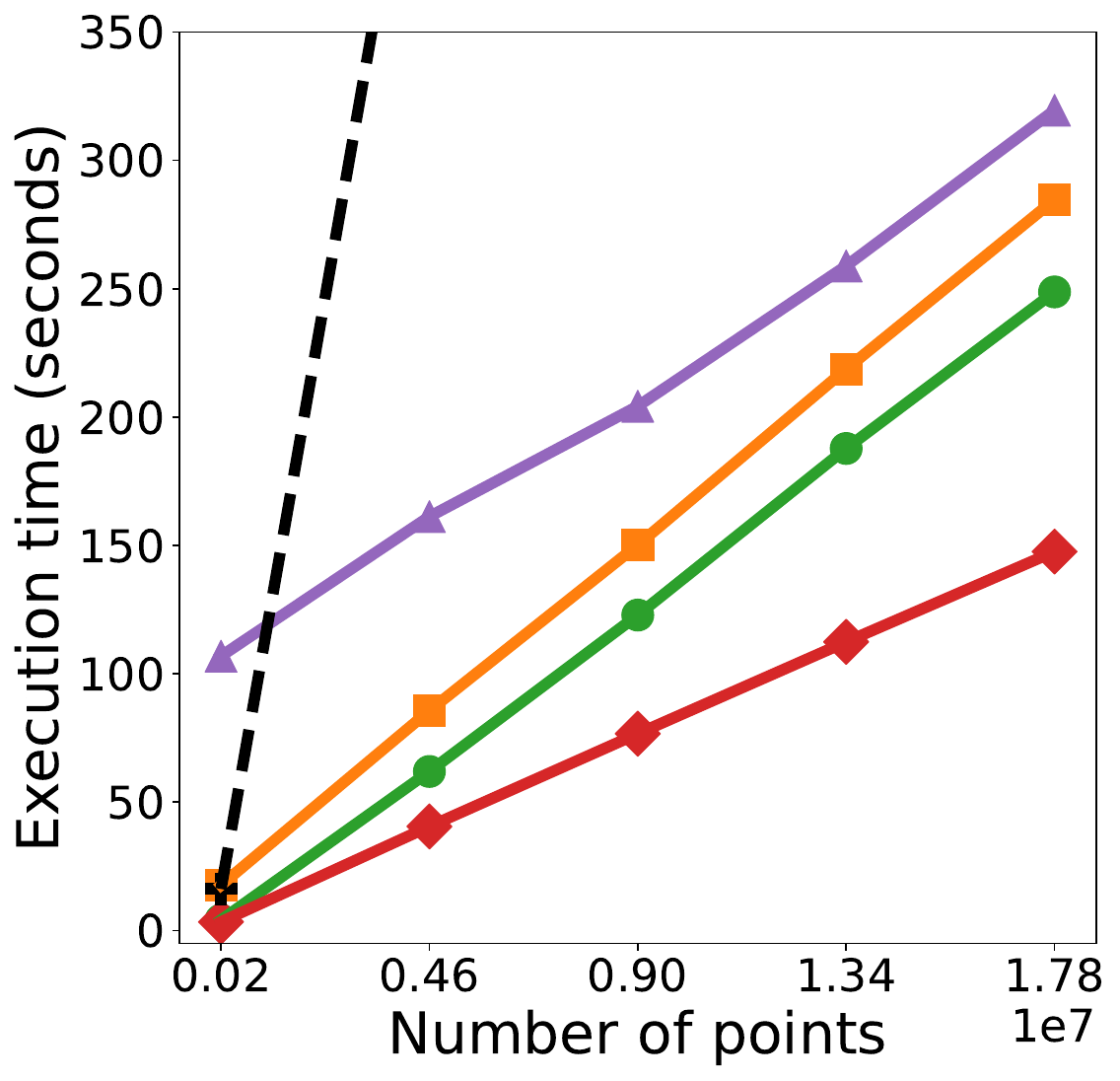} 
    \caption{\textit{k}-means clustering}
    \label{fig:comparison_plots_kmeans}
\end{subfigure} 
\hfill
\begin{subfigure}{0.328\textwidth} 
    \centering 
    \includegraphics[width=\linewidth,page=2]{Figures/Comparisons/KNN_comparison_shape.pdf} 
    \caption{KNN classification}
    \label{fig:comparison_plots_KNN} 
\end{subfigure} 
\hfill
\begin{subfigure}{0.315\textwidth} 
    \centering 
    \includegraphics[width=\linewidth,page=1]{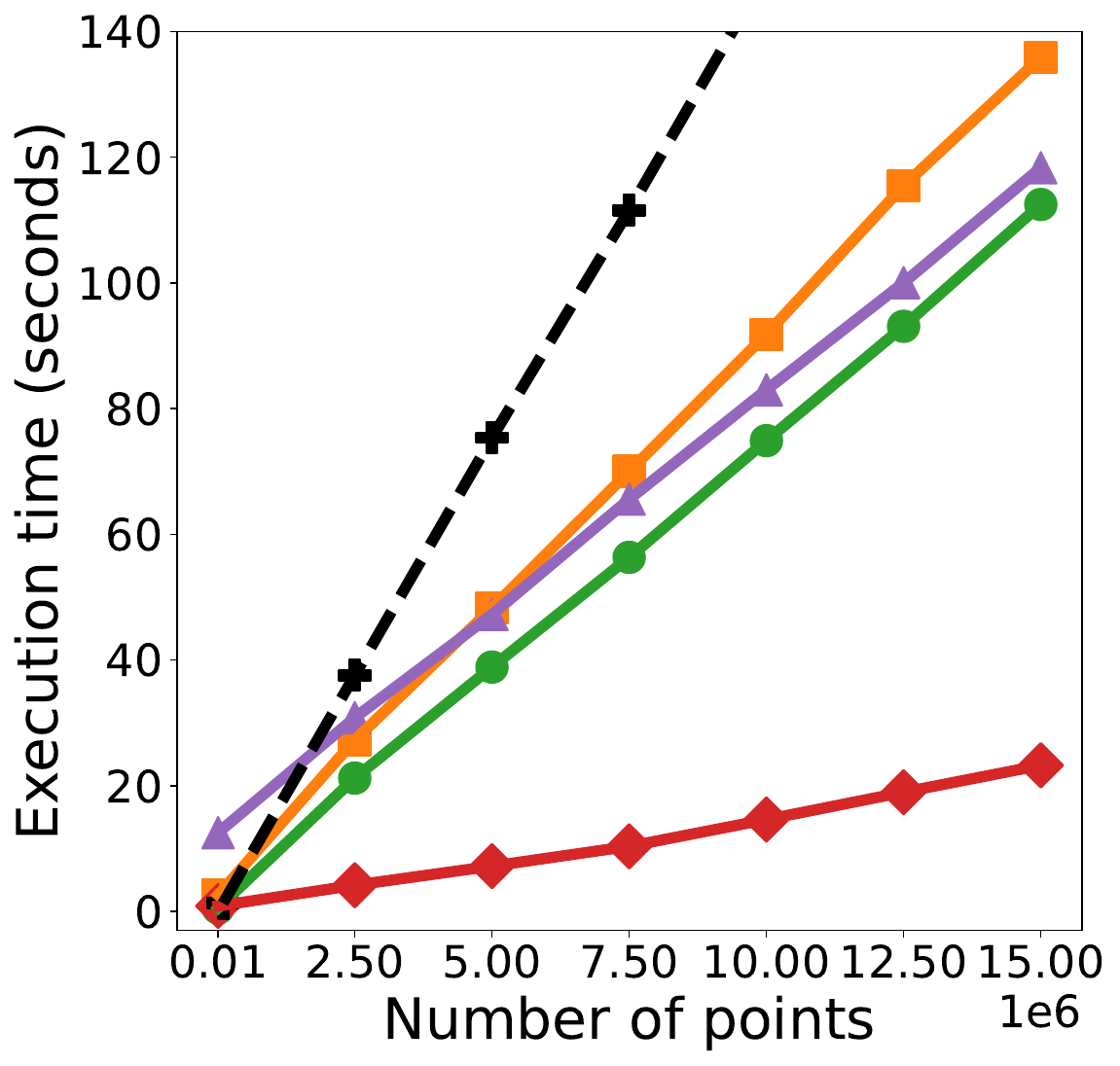} 
    \caption{Linear regression}
    \label{fig:comparison_plots_LR}
\end{subfigure}
\caption{Mean execution time (seconds) versus the number of points for \textit{k}-means, KNN, and linear regression on one multicore node. Methods are \textit{RCOMPSs}, \textit{future} and \textit{furrr} using \texttt{future\_multicore}, \textit{parallel} using \texttt{mclapply()}, and sequential R. The KNN panel has a logarithmic time axis.}
\label{fig:comparison_plots} 
\end{figure}

\javi{
For the single-node \textit{RCOMPSs} comparison, the runtime used one worker process for the node and one executor process per allocated core.
To contextualize these workflow differences, we compare execution-time scaling across three representative statistical workloads on a single-node multicore environment: an Intel Xeon Gold 6330 (Ice Lake) with 56 cores at 2.00\,GHz and 256\,GB of RAM. \xiran{In these experiments, the \textit{parallel} baseline refers to an implementation based on \texttt{mclapply()}. The \textit{future} and \textit{furrr} baselines used the same \texttt{future\_multicore} backend. For \textit{furrr}, chunking was not manually tuned; the implementation used \texttt{future\_map()} with the default scheduling/chunking behavior, with \texttt{furrr\_options()} used only to pass required packages or to disable automatic random-number seeding when explicit task-specific seeds were set.} \xiran{We selected these fork-based configurations as one set of single-node baselines to illustrate \textit{RCOMPSs} as a task-based option for the three workloads. Because \textit{RCOMPSs} does not rely on forked execution, this is not an execution-model-matched comparison. The reported comparison does not include \texttt{multisession}, \textit{mirai}, or other \textit{future} backends; consequently, the timings cannot establish a general performance ranking across R parallel frameworks. A matched non-forked benchmark would be needed for that purpose.} Figure~\ref{fig:comparison_plots} illustrates how mean execution time grows with workload size under the selected configurations. \sameh{\textit{future} and \textit{furrr} exhibit higher execution times at smaller workloads, while the relative difference decreases as the problem size increases.} \textit{parallel} scales more linearly but requires manual data partitioning. \textit{RCOMPSs} demonstrates consistent scaling across the three evaluated algorithms under the tested configurations. The trade-off is a baseline overhead that favors moderate-to-large workflows over trivially parallel prototypes. On distributed infrastructure, \textit{RCOMPSs} provides native multi-node execution through the \textit{COMPSs} runtime, while the future ecosystem can support distributed execution through additional backends and scheduler integrations.
}

\javi{
For \textit{k}-means clustering (Figure~\ref{fig:comparison_plots_kmeans}), \textit{future} has higher measured execution times at the smaller evaluated workloads, while \textit{RCOMPSs} shows lower growth in execution time over the evaluated problem sizes. These results describe the tested configurations; the present experiments do not isolate the mechanisms behind the observed differences. In KNN (Figure~\ref{fig:comparison_plots_KNN}), the synchronous distance calculations benefit \textit{parallel} and \textit{furrr} at smaller scales, but \textit{RCOMPSs} matches their efficiency at larger workloads while avoiding manual chunking configurations. For linear regression (Figure~\ref{fig:comparison_plots_LR}), \textit{RCOMPSs} exhibits a flatter execution-time growth curve over the evaluated problem sizes.
}


\javi{
It is important to note that \textit{RCOMPSs} provides a lower-level, runtime-oriented interface rather than a ready-to-use analytical API. It is intended to be wrapped by higher-level functions or pipeline frameworks that encapsulate task annotation, data routing, and error handling. This design preserves the simplicity of standard R syntax while exposing the underlying \textit{COMPSs} scheduling and dependency layers for developers building scalable statistical workflows. The practical trade-off is clear: \textit{RCOMPSs} introduces serialization and graph-construction overhead that may matter less as individual tasks become more computationally demanding, but can dominate short tasks. Users seeking transparent multi-node scalability and declarative dependency management will benefit from this paradigm; those prioritizing rapid iteration on modest hardware will find the modular \textit{future}-family more appropriate.
}

\xiran{A compact model-based summary of the Figure~\ref{fig:comparison_plots} trends, including fitted coefficients and goodness-of-fit values, is provided in the Supplementary Materials.}

\section{Conclusion}
\label{sec:Conclusions}

This work presented \textit{RCOMPSs}, a task-based programming model that bridges the R language with the \textit{COMPSs} framework to enable scalable parallel and distributed execution on multicore and manycore systems.
By integrating with the \textit{COMPSs} runtime, \textit{RCOMPSs} enables users to annotate standard R functions as tasks while \javi{leveraging the} dependency detection, scheduling, and data movement on \xiran{single-node and distributed-memory systems}.
\xiran{Existing R parallel frameworks are useful for rapid, in-memory prototyping through modular composition. \textit{RCOMPSs} takes a different approach: it leaves task orchestration, dependency management, and data movement to the \textit{COMPSs} runtime, trading initialization and serialization overhead for DAG-based execution and multi-node deployment.}
\xiran{This division keeps the structure of sequential R programs familiar, while developers still identify task boundaries and data dependencies.}
\xiran{Across the three workloads we evaluated, the results show how execution time scales under the tested multicore configurations and demonstrate distributed execution on HPC systems. These results illustrate practical workflow trade-offs rather than a universal performance advantage.}

Looking ahead, future work will focus on extending \textit{RCOMPSs} to GPU environments and enabling seamless integration with accelerator-based infrastructures. This will allow us to leverage \textit{COMPSs} support for GPU task offloading and benchmark against emerging GPU-enabled R frameworks, such as \textit{tensorflow} and CUDA-backed libraries. Additionally, we plan to explore adaptive scheduling strategies, fault tolerance mechanisms, and interoperability with modern R workflow systems. \xiran{Overall, \textit{RCOMPSs} gives R applications access to \textit{COMPSs} task execution while preserving R's expressiveness and leaving room for future support of heterogeneous systems.}


\section{Acknowledgment}
\label{sec:acknowledgment}

The authors gratefully acknowledge the support of the KAUST Supercomputing Laboratory (KSL) for providing access to the Shaheen~III supercomputer. Computational resources at the Barcelona Supercomputing Center (BSC) were provided through the MN5 system, and we thank the BSC Operations and Support teams for their assistance.


\section{Funding}
\label{sec:Funding}

This research was supported by the King Abdullah University of Science and Technology (KAUST) Research Funding (KRF) under Award No.\ ORFS-2022-CRG11-5069. Moreover, we acknowledge support from projects CEX2021-001148-S and PID2023-147979NB-C21, funded by MCIN/AEI and MICIU/AEI/10.13039/501100011033, and co-funded by the European Regional Development Fund (FEDER, EU). Additional support was provided by the Departament de Recerca i Universitats de la Generalitat de Catalunya through the research group MPiEDist (2021 SGR 00412).


\bibliographystyle{chicago}
\bibliography{JCGS/Bibliography}



\end{document}